\documentclass[XeLaTex,twocolumn,epjc3]{svjour3}          
\RequirePackage[T1]{fontenc}

\smartqed  

\RequirePackage{graphicx}
\RequirePackage{mathptmx}      
\RequirePackage{flushend}
\usepackage{amsmath}
\usepackage{amssymb}
\usepackage{epstopdf}
\RequirePackage[numbers,sort&compress]{natbib}
\RequirePackage[colorlinks,citecolor=blue,urlcolor=blue,linkcolor=blue]{hyperref}

\journalname{Eur. Phys. J. C}

\begin{document}

\title{Temporal evolution of a radiating star via Lie symmetries}

\author{Andronikos Paliathanasis\thanksref{e1,addr1,addr1b}
        \and Megandhren Govender\thanksref{e2,addr2}\and Genly Leon\thanksref{e3,addr3}
       }

\thankstext{e1}{e-mail: paliathanasis@gmail.com}
\thankstext{e2}{e-mail: megandhreng@dut.ac.za}
\thankstext{e3}{e-mail: genly.leon@ucn.cl}

\institute{Institute of Systems Science, Durban University of Technology, Durban, South Africa\label{addr1}\and Instituto de Ciencias F\'{\i}sicas y Matem\'{a}ticas, Universidad Austral de Chile, Valdivia 5090000, Chile\label{addr1b}\and Department of Mathematics, Faculty of Applied Sciences, Durban University of Technology, Durban, South Africa\label{addr2}\and Departamento de Matem\'{a}ticas, Universidad Cat\'{o}lica del Norte, Avda. Angamos 0610, Casilla 1280 Antofagasta, Chile.\label{addr3}}

\date{Received: date / Accepted: date}

\maketitle

\begin{abstract}
In this work we present for the first time the general solution of the temporal evolution equation arising from the matching of a conformally flat interior to the Vaidya solution. This problem was first articulated by Banerjee et al. (A. Banerjee, S. B. Dutta Choudhury, and Bidyut K. Bhui, {\em Phys. Rev.} D, {\bf 40} (670) 1989) in which they provided a particular solution of the temporal equation. This simple exact solution has been widely utilised in modeling dissipative collapse with the most notable result being prediction of the avoidance of the horizon as the collapse proceeds. We study the dynamics of dissipative collapse arising from the general solution obtained via the method of symmetries and of the singularity analysis. We show that the end-state of collapse for our model is significantly different from the widely used linear solution.
\end{abstract}

\section{Introduction}
The end state of a star resulting from continued gravitational
collapse is still a much debated topic in relativistic astrophysics.
It has been shown in some classes of models that shear plays an
important role in producing naked singularities. On the other hand,
the absence of shear during gravitational collapse of reasonable
matter distributions (for example in the case of perfect fluids)
always results in a black hole \cite{pankaj1}. An interesting study of gravitational collapse in the shear-free regime was proposed by Banerjee, Chaterjee and
Dadhich (hereafter referred to as the {\em BCD} model \cite{bcd}) in which the collapse proceeds without the formation of the horizon.
The so-called horizon-free collapse model has been explored in various contexts including higher dimensional spacetimes, Euclidean stars and class-one spacetimes \cite{hd1,euc1,k1,k2,k3}.

In this work we explore the boundary condition arising from the matching of a spherically symmetric imperfect fluid configuration undergoing dissipative collapse and matched to a Vaidya atmosphere. The boundary condition describing conservation of momentum flux across the bounding hypersurface is a second order nonlinear differential equation governing the temporal behaviour of the model. A particular solution of this equation which is linear in time has been widely used to model dissipative collapse in which the horizon never forms. We utilise a symmetry approach to generate the general solution of the boundary condition. We explore the physics associated with the collapsing core, particularly during the late stages of collapse.

This paper is structured as follows. In section $\S$\ref{sect:2} we present the field equations which govern the interior spacetime of the collapsing sphere and the junction conditions required for the smooth matching of the interior spacetime to the Vaidya exterior. We study the temperature profiles of the collapsing star in section $\S$\ref{sect:3}. A physical analysis of the thermodynamical variables and time of formation of the horizon is carried out in $\S$\ref{sect:4}. We conclude with an overall discussion in section $\S$\ref{sect:5}.
For completeness we add  \ref{app6}. In  \ref{app6.1}
is presented the stability analysis of powerlaw exact solutions. \ref{app6.2} is devoted to symmetries and singularity analysis.

\section{Dissipative collapse}
\label{sect:2}
When modeling a radiating star undergoing shear-free gravitational collapse the interior spacetime is described by the spherically symmetric line element \cite{bon1}
\begin{small}
\begin{equation} \label{1} ds^2 = -A^2(r,t) dt^2 + B^2(r,t)[dr^2 +
r^2 d \theta^2 + r^2 \sin^2 \theta d \phi^2]\,,
\end{equation} 
\end{small}
in which the metric functions $A$ and $B$ are yet to
be determined. The energy momentum tensor for the interior matter
distribution is described by an imperfect fluid given by \begin{equation} \label{2} T_{ab} = (\rho +
p) \, u_au_b + p g_{ab} + q_au_b + q_bu_a\,.
\end{equation} where $\rho$ and $p$ are the fluid energy density and pressure.
The heat flow vector $q^a$ is orthogonal to the
velocity vector so that $q^a u_a = 0$. The Einstein field
equations governing the interior of the stellar fluid is
\begin{eqnarray}   \label{t4}
 \rho &=& 3\frac{1}{A^2}\frac{{B_{t}}^2}{B^2} - \frac{1}{B^2}
\left( 2\frac{B_{rr}}{B} - \frac{{B_{r}}^2}{B^2} +
\frac{4}{r}\frac{B_{r}}{B} \right),  \label{t4a} \\ \nonumber \\
p &=& \frac{1}{A^2} \left(-2\frac{B_{tt}}{B} -
\frac{{B_{t}}^2}{B^2} +
2\frac{A_{t}}{A}\frac{B_{t}}{B} \right) \nonumber \\  \nonumber  \\
&& + \frac{1}{B^2} \left(\frac{{B_{r}}^2}{B^2} +
2\frac{A_{r}}{A}\frac{B_{r}}{B} + \frac{2}{r}\frac{A_{r}}{A} +
\frac{2}{r}\frac{B_{r}}{B} \right),  \label{t4b}  \\  \nonumber \\
p &=& -2\frac{1}{A^2}\frac{B_{tt}}{B} +
2\frac{A_{t}}{A^3}\frac{B_{t}}{B} -
\frac{1}{A^2}\frac{{B_{t}}^2}{B^2} +
\frac{1}{r}\frac{A_{r}}{A}\frac{1}{B^2} \nonumber \\ \nonumber \\
&& +  \frac{1}{r}\frac{B_{r}}{B^3} + \frac{A_{rr}}{A}\frac{1}{B^2} -
\frac{{B_{r}}^2}{B^4} + \frac{B_{rr}}{B^3},
\label{t4c}  \\ \nonumber \\
Q &=& -\frac{2}{AB} \left(-\frac{B_{rt}}{B} +
\frac{B_{r}B_{t}}{B^2} + \frac{A_{r}}{A}\frac{B_{t}}{B} \right),
\label{t4d} \end{eqnarray}
where $Q = \left(q_aq^a\right)^{1/2}$ is the magnitude of the heat flux.
We obtain the condition of pressure isotropy by equating
(\ref{t4b}) and (\ref{t4c}) \begin{equation}  \label{pi} 0 =
\frac{1}{B^2}\left[ \displaystyle\frac{A_{rr}}{A} +
\displaystyle\frac{B_{rr}}{B} - \left(2 \displaystyle\frac{B_r}{B} +
\displaystyle\frac{1}{r} \right) \left( \displaystyle\frac{A_r}{A} +
\displaystyle\frac{B_r}{B} \right)\right]. \end{equation}
This equation has been solved under various assumptions and transformations. It was Ivanov \cite{ive} who observed that if the constants of integration of a static solution to the pressure isotropy condition in comoving and isotropic coordinates are allowed to evolve with time, then, the time-dependent `solution' will automatically satisfy (\ref{pi}). Shear-free, radiating fluids in the presence of bulk viscosity were widely explored by Sussman. Starting with the most general spherically symmetric, shear-free metric, Sussman obtained a large family of exact solutions of the Einstein field equations which included Petrov type D, conformally flat and  self-similar solutions  \cite{sussman}. In particular, he showed that the constraints arising from the field equations for shear-free spacetimes can be treated as a linear algebraic system on the derivatives of the metric potentials.

Since the star is radiating
energy, the exterior spacetime is described by the
Vaidya metric \cite{v1}
\begin{small}
\begin{equation}
ds^2 = -\left(1-\frac{2m(v)}{\sf r}\right)dv^2 -
2dvd{\sf r} + {\sf r}^2[d \theta^2 + {\sf r}^2 \sin^2 \theta d
\phi^2] \label{vr}
\end{equation}
\end{small}
where $v$ is the retarded time and $m$ is the total mass inside
the comoving surface $\Sigma$ forming the boundary of the star. The
necessary junction conditions for the smooth matching of the
interior line element (\ref{1}) to the exterior spacetime (\ref{vr})
was first obtained by Santos \cite{santos} and has been succintly presented here for easy reference \begin{eqnarray}
(r B)_{\Sigma} &=& {\sf r}_{\Sigma}, \label{radius}\\
p_{\Sigma} &=& (q^1 B)_{\Sigma}, \label{p} \\
m_{\Sigma} &=& \Bigg[\frac{r^3 B {\dot B}^2}{2 A^2} - r^2 B' -
\frac{r^3 B'^2}{2 B} \Bigg]_{\Sigma}, \label{mass}\end{eqnarray} where
$m_{\Sigma}$ is the total mass within a sphere of radius
$r_{\Sigma}$ and (\ref{p}) represents the conservation of the
momentum flux across the boundary $\Sigma$.

An interesting approach to dissipative collapse is to explore the non-formation of the horizon in which the collapse rate is balanced by the rate at which energy is radiated to the exterior spacetime. The trio Banerjee, Chaterjee and Dadhich studied such a scenario by considering a simple radiative model in which the metric ansatz assumed is
\begin{eqnarray}
A &=& 1 + \zeta_0 r^2, \label{ss1} \\
B &=& R(t) \label{ss2}
\end{eqnarray}
where $\zeta_0$  and $C$ are positive constants. The
collapse evolves from $t = -\infty$ until $t=0$.
 Utilising the above
ansatz together with (\ref{t4b}) and (\ref{t4d}) in (\ref{p}) we
obtain
\begin{equation}
2R{\ddot R} + {\dot R}^2 + \alpha {\dot R} = \beta \label{bc}
\end{equation}
where $\alpha$ and $\beta$ are constants.
A special and simple solution to this equation is $R = -Ct$ where $C>0$ is a constant. Since the star is collapsing, we require the expansion scalar, $\Theta = \frac{3}{A}\frac{\dot R}{R} < 0$. This solution first made its appearance in the literature in 1989 when Banerjee et al \cite{bhui} presented the most general class of conformally flat radiating solutions. While the solution has a simple form it is remarkable that it has revealed rich and diversified toy models of dissipative collapse. It has been  observed in the {\em BCD} paper that the ratio
\begin{equation}
1 - \frac{2m_\Sigma}{r_\Sigma}
\end{equation}
is independent of time. This can be easily seen from equations (\ref{radius}) and (\ref{mass}) in which both the area radius and mass are in linear in $t$. Thus the ratio of mass to area radius is independent of time and the boundary surface cannot reach the horizon.

\subsection{A new radiating model}

We now model a radiating star undergoing dissipative collapse by digressing from the simple linear time-dependence of $B(r,t)$. To this end we utilise the truncated solution reported in $\S$7.2 and derived by using the Lie symmetry analysis,
\begin{equation} \label{tbc}
R(t) = R_0t^{2/3} - \frac{3\alpha}{4}t\end{equation}
with $\beta = -\frac{3\alpha^2}{4}$. 
The Einstein field equations (\ref{t4a})--(\ref{t4d})
reduce to
\begin{small}
\begin{eqnarray}
\rho &=& \frac{\left(8R_0 - 9\alpha t^{1/3}\right)^2}{3t^2\left(-4R_0 + 3\alpha t^{1/3}\right)^2\left(1 + \zeta_0 r^2\right)^2}, \label{newrho}\\
p&=& \frac{32\alpha R_0 + 3t^{1/3}\left(-9\alpha^2 + 64\zeta_0(1+ r^2\zeta_0)\right)}{3t^{5/3}\left(4R_0 - 3\alpha t^{1/3}\right)^2\left(1 + \zeta_0r^2\right)^2}, \label{newp}\\
qB&=&-\frac{55\left(-32\alpha R_0 + 27\alpha^2t^{1/3} - 192\zeta_0t^{1/3}(1 + \zeta_0r^2)\right)}{3t^{5/3}\left(-4R_0 + 3\alpha t^{1/3}\right)^2\left(1+ \zeta_0r^2\right)^2}.
\end{eqnarray}
\end{small}
We also calculate the mass function and luminosity at infinity to be
\begin{eqnarray}
m &=& \frac{r^2\left(8R_0 - 9\alpha t^{1/3}\right)^2\left(4R_0 - 3\alpha t^{1/3}\right)}{1152\left(1 + \zeta_0r^2\right)^2}, \label{newmass}\\
L_\infty &=& \frac{\alpha r_0\left(32R_0 - 27\alpha t^{1/3}\right)\left(8R_0 - 9\alpha t^{1/3}\right)\Xi}{864t^{7/3}\left(4R_0 - 3\alpha t^{1/3}\right)^2\left(1 + \zeta_0 r_0^2\right)^4} \label{lu}
\end{eqnarray}
where we have defined
\begin{equation}
\Xi = 8R_0r_0 + 3t^{1/3}(4 - 3\alpha r_0 + 4\zeta_0r_0^2)
\end{equation} and
$r = r_0$ defines the boundary of the star at some fixed time.

\subsection{The role of an equation of state}

There have been many attempts at incorporating an equation of state (EoS) in radiating stellar models. This is an important requirement which relates the pressure to the fluid density and this in turn makes the interplay between these thermodynamical quantities transparent. An early attempt at imposing a linear equation of state of the form, $p_r = \gamma \rho$ where $\gamma$ is a constant was carried out by Wagh et al. \cite{wagh1} in which they matched a shear-free interior metric to the Vaidya exterior. In their model the temporal dependence was linear. More recently, Bogadi et al. \cite{rs1} attempted to model a radiating star which collapses from an initial static configuration. In this work, both the static and dynamical models obey a linear EoS. In order to satisfy the boundary condition, it was shown that the EoS parameter, $\gamma = -1/3$. This investigation demonstrated the difficulty of imposing an EoS for the dynamical model, i. e., the gravitational potentials must simultaneously satisfy the EoS at each interior point $(r,t)$ of the collapsing body and the boundary condition (where $r = r_0$, is constant). At best, one can choose an initial static configuration which satisfies a given linear EoS. This choice automatically changes the dynamical boundary condition as the radial pressure vanishes at the boundary of the static configuration. In our work, the metric potentials $(A,B)$ given by (\ref{ss1}) and (\ref{ss2}) {\em do not} describe an initial static configuration. At the very least the simple forms chosen for $(A,B)$ here could mimic an EoS in the appropriate limit but this carries no guarantee until the boundary condition is satisfied. In the BCD model the energy density and pressure both diverge as $1/t^2$ and the mass evolves linearly with time. In addition, the ratio $p/\rho$ is independent of time and increases as $r^2$. In our model $p/\rho$ increases as times evolves and changes as $1/r^3$. It is clear that relationship between pressure and density is significantly altered when higher order terms are included in the temporal behaviour of the model.

\section{Causal heat flow}
\label{sect:3}
The role of causal heat flow during dissipative collapse has been extensively studied by Herrera and co-workers \cite{hh1,hh2,hh3} and references therein. It has been demonstrated that relaxational effects lead to higher core temperatures as the collapse proceeds with cooling being enhanced in the surface layers. The noncausal Eckart formalism may hold during an epoch when the fluid is close to hydrostatic equilibrium. However, as the collapse proceeds the noncausal nature of the Eckart framework leads to infinite propagation speeds of the thermal signals and unstable equilibrium states. Earlier work by Di Prisco et al. \cite{hh1} has shown that relaxational effects impact on the luminosity profiles of radiating stars.
In order to study the impact of relaxation times on the temperature profiles we adopt a causal heat transport equation of Maxwell-Cattaneo form \cite{mg1}
\begin{equation}
	\tau {h_a}^b {\dot{q}}_b + q_a = - \kappa (D_a T +
	T {\dot{u}}_a), \label{r2.26}
\end{equation} where the relaxation time is given by
\begin{equation} \label{r2.24}
	\tau = \kappa T \beta
\end{equation} for the heat flux.
The appearance of the relaxation time restores causality and has been successful in modelling high-frequency phenomena in electronics and fluid flow \cite{royy}. For the line element (\ref{1}) the causal heat transport
equation (\ref{r2.26}) becomes \begin{equation} \label{ca1}
	\tau(qB)_{,t} + A(qB) = -\kappa \frac{(AT)_{,r}}{B},
\end{equation} which governs the behavior of the temperature.
Setting $\tau = 0$ in (\ref{ca1}) we obtain the familiar Fourier
heat transport equation \begin{equation} \label{ca2} A(qB) =
	-\kappa \frac{(AT)_{,r}}{B}, \end{equation} which predicts
reasonable temperatures when the fluid is close to
quasi--stationary equilibrium.

Following the work of \cite{mg2} we adopt the following thermodynamic coefficients for radiative transfer where we assume that heat is being carried away from the core via thermally generated neutrinos. The thermal
conductivity assumes the form \begin{equation} \kappa =\chi
	T^3{\tau}_{\mathrm c} \label{a28}\,,\end{equation} where $\chi$ ($\geq0$) is a constant and ${\tau}_{\mathrm c}$ represents the mean
collision time between massless and massive particles. Martinez \cite{mart} has shown that $\tau_{\mathrm c} \propto T^{-3/2}$ for thermally generated neutrinos within the core of neutron stars. To this end we assume
\begin{equation} \label{a29} \tau_{\mathrm c}
	=\left(\frac{\psi}{\chi}\right) T^{-\omega} \,,\end{equation}
where $\psi$ ($\geq 0$) and $\omega$ ($\geq 0$) are constants and for  $\omega=\frac{3}{2}$ we regain the treatment due to Martinez. We observe that this form implies that the mean collision time decreases with an increase in temperature, as expected. We assume that relaxation time is proportional to the
collision time: \begin{equation} \tau =\left(\frac{\lambda \chi }{\psi}\right) \tau_{\mathrm c} \label{a30}\,,\end{equation} where
$\tau$ ($\geq 0$) is a constant.
This assumption may hold for a limited epoch of the collapse process.
Putting all together in (\ref{ca1}) we obtain \begin{equation} \lambda (qB)_{,t} T^{-\omega} + A (q
	B) = - \psi \frac{T^{3-\omega} (AT)_{,r}}{B} \label{temp1}
	\,.\end{equation}
It has been shown that in the case $\omega=0$ (which corresponds to constant mean collision time), the causal transport equation (\ref{temp1}) yields the following temperature profile \cite{kesh1}
\begin{equation} (AT)^4 = - \frac{4}{\psi} \left[\lambda\int A^3 B
	(qB)_{,t}{\mathrm d} r + \int A^4 q B^2 {\mathrm d} r\right] +
	{\cal{F}}(t) \label{caus0} \end{equation} where ${\cal{F}}(t)$ is an integration function.
	We can evaluate ${\cal{F}}(t)$ by recalling that the effective surface temperature of a star is given by
\begin{equation}
\left(T^4\right)_\Sigma = \left(\frac{1}{r^2B^2}\right)_\Sigma\left(\frac{L_\infty}{4\pi \sigma}\right)_\Sigma
\end{equation}
where $\delta (> 0)$ is a constant. For our model, we are able to complete the integration in (\ref{caus0}) and the temperature is plotted in figure 6.

\section{Physical analysis}
\label{sect:4}
\begin{figure}[!ht]
 	\centering
 	\includegraphics[scale=0.6]{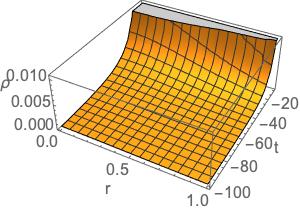}
 	\caption {Density as a function of the radial and temporal coordinates}
 	\label{Fig:Density}
 \end{figure}

\begin{figure}[!ht]
 	\centering
 	\includegraphics[scale=0.6]{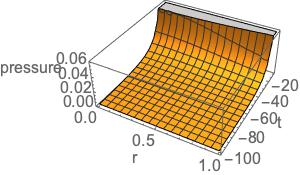}
 	\caption {Pressure as a function of the radial and temporal coordinates}
 	\label{Fig:p}
 \end{figure}

\begin{figure}[!ht]
 	\centering
 	\includegraphics[scale=0.6]{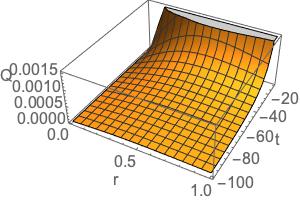}
 	\caption {Heat flow as a function of the radial and temporal coordinates}
 	\label{Fig:Q}
 \end{figure}

\begin{figure}[!ht]
 	\centering
 	\includegraphics[scale=0.6]{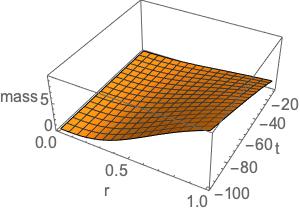}
 	\caption {Mass profile as a function of the radial and temporal coordinates}
 	\label{Fig:mass}
 \end{figure}

\begin{figure}[!ht]
 	\centering
 	\includegraphics[scale=0.6]{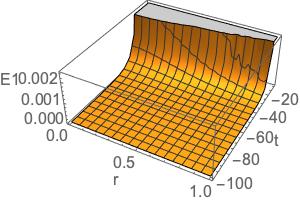}
 	\caption {$E1 = (\rho + p)^2 - 4Q^2 > 0$ as a function of the radial and temporal coordinates}
 	\label{Fig:E1}
 \end{figure}

\begin{figure}[!ht]
 	\centering
 	\includegraphics[scale=0.6]{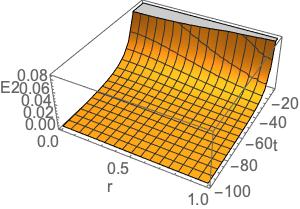}
 	\caption {$E2 = \rho - 3p + \left[(\rho + p)^2 - 4Q^2\right]^{1/2} > 0$ as a function of the radial and temporal coordinates}
 	\label{Fig:E2}\end{figure}

\begin{figure}[!ht]
 	\centering
 	\includegraphics[scale=0.6]{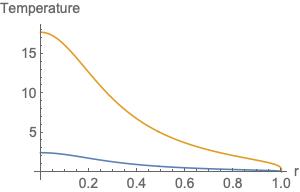}
 	\caption {Temperature profiles as a function of the radial coordinate}
 	\label{Fig:t}\end{figure}

In order to establish the physical viability of our model, we have plotted the density, pressure, heat flow and the energy conditions, respectively as functions of the radial and temporal coordinates in Figs. \ref{Fig:Density}, \ref{Fig:p}, \ref{Fig:Q}, \ref{Fig:mass}, \ref{Fig:E1} and \ref{Fig:E2}. We observe that the density (Fig. \ref{Fig:Density}) and pressure (Fig. \ref{Fig:p}) are monotonically decreasing functions of the radial coordinate. Furthermore, as the collapse proceeds, the density and pressure increase. This is expected as the core collapses the matter gets squeezed into smaller volumes thus increasing the density and the pressure within the smaller sphere. In the case of the BCD model, the rate at which the body radiates energy is balanced by the rate at which it collapses which leads to the final `evaporation' of the star. It is worth pointing out that the luminosity as given by (\ref{lu}) vanishes when
\begin{eqnarray}
t_{(bh)_1} &=& 0.702332\nu^3,\\
t_{(bh)_2} &=& 1.66479 \nu^3,\\
t_{(bh)_3} &=& -18.963\frac{\alpha^3\nu^3}{(4 - 3r_0\alpha +4r_0^2\zeta_0)^3},
\end{eqnarray}
where we have defined $\nu = \frac{R_0}{\alpha}$. Since the collapse proceeds from $-\infty < t < 0$, we must have $t_{(bh)_i} < 0$ if we want the horizon to form before the collapse ends. For $\alpha > 0$ we must $R_0 < 0$ which ensures that $t_{(bh)_1}$ and $t_{(bh)_2}$ are both negative. For $t_{(bh)_3} < 0$ we must have $4 - 3r_0\alpha + 4r_0^2\zeta_0$ which places a restriction on $\zeta_0$. In the BCD model where the horizon never forms, no such restriction is placed on $\zeta_0$.
Fig. \ref{Fig:Q} shows that the heat flux (energy output) increases with time. As the density increases, the thermonuclear processes become more efficient resulting in higher energy outputs for late times. The mass profile is depicted in Fig. \ref{Fig:mass}. The mass of the sphere increases monotonically from the center to the boundary. The mass decreases with time as star radiates energy as it collapses. Figs. \ref{Fig:E1} and \ref{Fig:E2} confirm that the energy conditions required for a realistic stellar core are obeyed everywhere inside the star. We have plotted the causal temperature profiles in Fig. \ref{Fig:t}. The blue curve represents the BCD temperature profile whereas the orange curve describes the evolution of the temperature as a function of the radial coordinate for our nonlinear solution. It is evident that the nonlinear solution predicts higher core temperatures than the linear BCD model. At this juncture it is necessary to point out that the condition of pressure isotropy which we employed to generate the static solution is unstable in the sense that such an isotropic matter distribution will evolve into an anisotropic regime as it leaves hydrostatic equilibrium.  The presence of dissipative fluxes (in the form of heat flow), density inhomogeneities and/or nonzero shear within the fluid flow, can disrupt the condition of pressure isotropy leading to unequal radial and tangential stresses within the collapsing fluid \cite{pi}.
Our aim of this work was to demonstrate the existence of the general solution of the junction condition which encodes the temporal behaviour of the model. We observe that the dynamical behaviour (temperature profiles and horizon formation) are clearly different from the simple linear temporal evolution assumed in the BCD model. We have shown that the inclusion of higher order terms in the temporal evolution of the model significantly affects the dynamics of the collapse process. This can be seen in the altered temperature distribution at each interior point of the collapsing fluid and the time of formation of the horizon. 

\section{Concluding remarks}
\label{sect:5}
In this exposition we analysed a differential equation arising from the modeling of a star undergoing dissipative collapse. We have presented the general solution to the boundary condition for a particular type of shear-free, dissipative collapse. In this model the gravitational potentials are separable in the radial and temporal coordinates and the pressure is isotropic at each interior point of the collapsing distribution. We carried out an extensive stability analysis of the solution arising from the temporal equation in \ref{app6.1}.  We have analytically proved the following results: (i)
The solution of \eqref{bc}
$R = -Ct$ defined for $-\infty < t \leq 0$, where $C>0$ is a fixed constant by Banerjee
et al. \cite{bhui}, is unstable as $t\rightarrow 0^-$ and stable as $t\rightarrow -\infty$. (ii) We found a family of solutions  of \eqref{bc} given by
\begin{equation}
\label{eq81}
R(t)=   \frac{\alpha  c_1}{2-\theta^*}+\frac{\alpha  (\theta^*-1)    t}{2-\theta^*}, \quad c_1\neq 0,
\end{equation}
parametrised by
$-\infty<\theta^*<\infty$. They are defined in the semi-infinite-interval
$-\infty< t\leq 0$,
and are stable as $t\rightarrow 0^{-}$. Our solutions are intrinsically different from the closed-form solution of Banerjee
et al. [1] in terms of their stability as well as in nature, given that $\lim_{t\rightarrow 0^{-}}  R(t) = \frac{\alpha  c_1}{2-\theta^*} \neq 0$, 
as a difference with the closed-form solution of Banerjee
et al. \cite{bhui} which satisfies $ \lim_{t\rightarrow 0^{-}}  R(t) =0$. The affine parameter $\frac{\alpha  c_1}{2-\theta^*}$ makes a subtle difference concerning stability as $t\rightarrow 0^{-}$.
 We showed that a particular nonlinear temporal dependence produces drastically different physics from the linear model. This is an important point to note, albeit that the collapsing sphere described here represents a toy model of dissipative collapse.
 
 Our model does not feature an EoS 
as the chosen metric functions are simplistic in nature. An inclusion of an EoS will form the basis of future work.

\section*{Acknowledgments}

AP \& GL were funded by Agencia Nacional de Investigaci\'on y Desarrollo - ANID through the program FONDECYT Iniciaci\'on grant no. 11180126. Additionally, GL thanks the support of Vicerrector\'ia de Investigaci\'on y Desarrollo Tecnol\'ogico at Universidad Cat\'{o}lica del Norte. We thank the anonymous referee for his/her valuable comments which have helped us  improve our work. 

\appendix

\section{Dynamical systems analysis, symmetries and singularity analysis}
\label{app6}

\subsection{Stability analysis of exact solutions}

\label{app6.1} We have the main equation \eqref{bc} where $\alpha$ and $%
\beta $ are constants. For convenience we assume $\alpha>0$. A special and
simple solution to this equation is $R = -Ct$ defined for $-\infty < t \leq
0 $, where $C>0$ is a constant given by the positive root of $-\beta
+C^2-\alpha C=0$.

For analysis of the stability of the scaling solution $R_{s}(t)=-Ct$, with $%
C=\frac{1}{2}\left( \sqrt{\alpha ^{2}+4\beta }+\alpha \right) >0$ in the
interval $-\infty <t<0$ we use similar methods as in Liddle \& Scherrer
\cite{Liddle:1998xm} and in Uzan \cite{Uzan:1999ch}. Defining the new time
variable $\eta $ through $t=-e^{-\eta },-\infty <\eta <\infty$ such that $%
t\rightarrow -\infty $ as $\eta \rightarrow -\infty $ and $t\rightarrow 0$
as $\eta \rightarrow +\infty $, as well as the variables $\epsilon (\eta )=%
\frac{R(\eta )}{R_{s}(\eta )}-1,\quad v(\eta )=\epsilon ^{\prime }(\eta
)$ in which $R(\eta )=R(-e^{-\eta })$ and $R_{s}(\eta )=-Ct(\eta
)=Ce^{-\eta }$ allows the scaling solution to be translated to the origin.
Finally, we obtain the system {\small \
\begin{eqnarray}
&&\epsilon ^{\prime }=v, \\
&&v^{\prime }=\frac{\epsilon (\alpha -C\epsilon -2C)}{2C(\epsilon +1)}%
+v\left( 2-\frac{\alpha }{2C\epsilon +2C}\right) -\frac{v^{2}}{2(\epsilon +1)%
},
\end{eqnarray}%
} with linearisation matrix $J(\varepsilon ,v)$. The latter system admits
the stationary points $P_{1}=\left( 0,0\right) $ and $P_{2}=\left( \frac{%
\alpha }{C}-2,0\right) $. At point $P_{1}$, the matrix $J(0,0)$ has
eigenvalues $\left\{ 1,1-\frac{\alpha }{2C}\right\} $. Assume first $\beta
\geq 0$. In this case, the origin is always unstable as $\eta \rightarrow
+\infty $ due to $2C=\alpha +\sqrt{\alpha ^{2}+4\beta }>\alpha $. The origin
is stable as $\eta \rightarrow -\infty $. Moreover, at the point $P_{2}$ we
find that the matrix $J(\frac{\alpha }{C}-2,0)$ has eigenvalues $\left\{ 1,\frac{%
2C-\alpha }{2(C-\alpha )}\right\} $. Due to $2C-\alpha >0$ it is unstable as
$\eta \rightarrow \infty $. Indeed for $0<\frac{\alpha }{2}<C<\alpha $ it is
a saddle, whereas for $C>\alpha >0$ is an unstable node. The last condition
is forbidden due to the physical condition $\epsilon \geq -1$ evaluated at $%
(\epsilon ,v)=(\frac{\alpha }{C}-2,0)$ implies $\alpha \geq C$.~In the case $%
\beta <0,~\alpha <-2\sqrt{-\beta }$ or $\beta <0,\alpha >2\sqrt{-\beta }$%
,~we have two solutions $R_{s\pm }(\eta )=-C_{\pm }t(\eta )=C_{\pm }e^{-\eta
},$ where $2C_{\pm }=\alpha \pm \sqrt{\alpha ^{2}-4|\beta |}$. Observe that $%
2C_{+}>\alpha $, implies that $R_{s+}(\eta )$ is an unstable solution
(unstable node) as $\eta \rightarrow \infty $. Due to $\alpha >2C_{-}>0$, $%
R_{s-}(\eta )$ is an unstable (saddle) solution.

On the other hand, there are non-trivial dynamics as $(\epsilon ,v)$ are
unbounded. Let us now define $\phi :=h(\epsilon )=(\epsilon +1)^{-1/n}~$, $%
\theta =v-\epsilon ,$ where the coordinate transformation $\phi =h(\epsilon )$
for $n>1$ maps the interval $[\epsilon _{0},\infty )$ onto $(0,\delta ]$,
with $\delta =h(\epsilon _{0})$, satisfying $\lim_{\epsilon \rightarrow
+\infty }h(\epsilon )=0$. We then obtain the dynamical system
\begin{eqnarray}
&&\phi ^{\prime }=-\frac{\phi }{n}+\left( \frac{1}{n}-\frac{\theta }{n}%
\right) \phi ^{n+1},  \label{eq30} \\
&&\theta ^{\prime }=\phi ^{n}\left( \left( 1-\frac{\alpha }{2C}\right)
\theta -\frac{\theta ^{2}}{2}\right) .  \label{eq31}
\end{eqnarray}%
where $2C-\alpha >0$ and$~0<\phi <\delta ^{1/2}$,~$\theta \in K,$ where $K$
is a compact set. Neglecting higher order terms $\propto\phi ^{n+1}$ as $\phi\rightarrow 0$ in Eq. (\ref{eq30}) we obtain the asymptotic solution:~$\phi (\eta )=e^{-\frac{\eta }{n}}c_{1},$ $~\theta
(\eta )=\frac{\left( 2C-\alpha \right) }{C-\exp \left( \frac{(2C-\alpha
)\left( e^{-\eta }c_{1}^{n}+2Cc_{2}\right) }{2C}\right) }.$
The system (\ref{eq30}), (\ref{eq31}) admits a line of
fixed points $L:(\phi ,\theta )=(0,\theta ^{\ast })$ as $\eta \rightarrow
\infty $ for bounded $\theta $. As $\eta \rightarrow \infty $ the
eigenvalues associated to $L$ are $\left\{ -\frac{1}{n},0\right\} $.
Therefore, it is normally hyperbolic and stable. It is verified that asymptotic solution showed before converges to it.

A summary of our analytical findings indicate: (i) The solution of (%
\ref{bc}) $R=-Ct$ defined for $-\infty <t\leq 0$, where $C>0$ is a fixed
constant, is unstable as $t\rightarrow 0^{-}$ and stable as $t\rightarrow
-\infty $. (ii) The curve of fixed points $L:(\phi ,\theta )=(0,\theta
^{\ast })$ (i.e., $\epsilon \rightarrow \infty $, and $v\rightarrow \infty $%
, in such a way that $v-\epsilon\rightarrow \theta ^{\ast }$) is stable as
$\eta \rightarrow \infty $ for bounded $\theta $. Result (ii) means that, as
$\eta \rightarrow \infty $, we have $\epsilon ^{\prime}(\eta )=\epsilon (\eta )$,$~\epsilon (\infty )=\infty .$ Hence, $\epsilon (\eta )=c_{1}e^{\eta
}-\theta ^{\ast }$ ,$~c_{1}\neq 0.$ Then,~$R(\eta )=C\left( c_{1}-(\theta
^{\ast }-1)e^{-\eta }\right) $. 

Substituting into (\ref{bc}), we find the
constraint $C\theta ^{\ast }(\alpha +C(\theta ^{\ast }-2))=0.$ We have some
specific solutions when $\theta ^{\ast }\in \left\{ 0,2-\frac{\alpha }{C}%
\right\} $. Recall that $\theta ^{\ast }$ is an arbitrary constant
value by definition of line $L$, so the natural condition is $C=\frac{%
\alpha }{2-\theta ^{\ast }}.$ Then, the solution of (\ref{bc}) given by %
\eqref{eq81} defined in the semi-infinite-interval $-\infty <t\leq 0$, is
stable as $t\rightarrow 0^{-}$ ($\eta \rightarrow +\infty $). Finally, $%
\lim_{t\rightarrow 0^{-}}R(t)=\frac{\alpha c_{1}}{2-\theta ^{\ast }}\neq 0~$%
by construction. We should point out that the in(stability) of the solutions demonstrated in this section is not related to the instability of the pressure isotropy condition highlighted in \cite{pi}. 

\subsection{Symmetries and singularity analysis}

\label{app6.2}

Let us now turn our attention to the boundary condition (\ref{bc}). Equation
(\ref{bc}) is invariant under the infinitesimal transformation \cite{kumei} $%
\bar{t}\rightarrow t+\varepsilon \left( \alpha _{1}+\alpha _{2}t\right) $,~$%
\bar{R}\rightarrow R+\varepsilon \left( \alpha _{2}R\right) \,\ $where $%
\varepsilon $ is an infinitesimal parameter.$~$We infer that equation (\ref%
{bc}) admits as Lie point symmetries the elements of the two-dimensional Lie
algebra $\left\{ \partial _{t},t\partial _{t}+R\partial _{R}\right\} $ which
form the $A_{2,2}$ Lie algebra in the Morozov-Mubarakzyanov classification
scheme \cite{mb1,mb2,mb3,mb4}. The application of the Lie symmetries in
gravitational physics has provided us with many interesting results, for
instance see \cite{sym1,sym2,sym3,sym4,sym5} and references therein.

\bigskip From the vector field $\partial _{t}$ we find the reduced equation $%
2xy\frac{dy\left( x\right) }{dx}+y^{2}\left( x\right) +\alpha y\left(
x\right) -\beta =0~$ with $y\left( x\right) =\dot{R}~,~x=R$. The reduced
equation gives
\begin{eqnarray}
\ln \left( x-x_{0}\right)  &=&-\ln \left( y^{2}+\alpha y-\beta \right) + \\
&&-\frac{2\alpha }{\sqrt{\alpha ^{2}+4\beta }}\text{arctanh}\left( \frac{%
2y+\alpha }{\sqrt{\alpha ^{2}+4\beta }}\right) .
\end{eqnarray}%
or in the special case where $\beta =-\frac{\alpha ^{2}}{4}$, the reduced equation
provides $\ln \left( 2+y+\alpha \right) +\frac{\alpha }{2y+\alpha }=-\frac{1}{2}%
\ln \left( x-x_{0}\right) .$

On the other hand, from the Lie symmetry $t\partial _{t}+R\partial _{R}$ for
equation (\ref{bc}) it follows $2z\left( y\left( z\right) -z\right) \frac{%
dy\left( z\right) }{dz}+y^{2}\left( z\right) +\alpha y\left( z\right) -\beta
=0.~$This is an Abel equation and can be integrated by using a Lie
integration factor. A special solution is the constant value of $y=y_{0}$
with $y_{0}^{2}+\alpha y_{0}-\beta =0$, however, such solution leads to the
closed-form solution of Banerjee et al. \cite{bhui}.

We proceed with our analysis by writing a closed form solution of equation (\ref%
{bc}) derived with the singularity analysis. The modern treatment of the
singularity analysis is described by the ARS algorithm \cite%
{ars1,ars2,ars3,anl1}. For equation (\ref{bc}) \ the leading-order term is
found to be $R_{leading}\left( t\right) =R_{0}\left( t-t_{0}\right) ^{\frac{2%
}{3}}$, where $t_{0}$ indicates the location of the movable singularity and $%
R_{0}$ is arbitrary. The resonances, are derived to be $s_{1}=-1$ and $%
s_{4}=4$, which means that the analytic solution of (\ref{bc}) can be
expressed in terms of the Right Painlev\'{e} Series%
\begin{equation}
R\left( t\right) =R_{0}\left( t-t_{0}\right) ^{\frac{2}{3}%
}+R_{1}t+R_{2}\left( t-t_{0}\right) ^{\frac{4}{3}}+R_{3}\left(
t-t_{0}\right) ^{\frac{5}{3}}+...~.
\end{equation}%
We replace in (\ref{bc}) from where we find that $R_{1}=-\frac{3\alpha }{4}%
,~R_{2}=\frac{9}{320R_{0}}\left( 3\alpha ^{2}+16\beta \right) ~,~R_{2}=\frac{%
3\alpha }{320R_{0}^{2}}\left( 3\alpha ^{2}+16\beta \right) ,...~$.~However,
in the special case in which $\left( 3\alpha ^{2}+16\beta \right) =0$, that
is $\beta =-\frac{3\alpha ^{2}}{16}$, we find that $R_{I}=0,~I>1$.


\begin{thebibliography}{}
	
	
		\bibitem{bhui} A. Banerjee, S.B. Duttachoudhury and B.K. Bhui, {\em Phys. Rev} {\bf D40}, 670
		(1989).
		
		\bibitem{8} S.R. Maiti, {\em Phys. Rev.} {\bf D25}, 2518 (1982).
		
		\bibitem{pankaj1} K. Mosani, D. Dey and P. S. Joshi, {\em Phys. Rev.} D, 	{\bf 101}  044052 (2020).
		
	
	\bibitem{bcd} A. Banerjee, S. Chatterjee and N. Dadhich, {\em Mod.
		Phys. Lett. A} {\bf 35}, 2335 (2002).
	
	\bibitem{hd1} A. Banerjee and S. Chaterjee, {\em Astrophys. Space Sci.}, {\bf 299} 3 (2004).
	
	\bibitem{euc1} K. S. Govinder and M. Govender, {\em Gen. Relativ. Grav.}, {\bf 44} 147 (2012).
	
	\bibitem{k1}
	N.F. Naidu, M. Govender, S.D. Maharaj, {\em Eur. Phys. J.} C. {\bf  78}, 48 (2018)
	
	\bibitem{k2}
	J. Ospino, L.A. Nunez, {\em Eur. Phys. J.} C. {\bf  80}, 166 (2020)
	
	
	\bibitem{k3}
	S.C. Jaryal, {em Eur. Phys. J.} C. {\bf  80}, 683 (2020)
	
		\bibitem{bon1}  W.B. Bonnor, A.K. de Oliveira \& N.O. Santos, {\em Phys. Rep.} {\bf 181}, 269
		(1989).
		
	\bibitem{ive} B. V. Ivanov,  {\em Gen. Relativ. Gravit.} {\bf 44}, 1835 (2012).
	
	\bibitem{sussman} R. A. Sussman,  {\em Class. Quantum Grav.} {\bf 10},  2675 (1993).
		
	\bibitem{v1}  P.C. Vaidya, {\em Proc. Ind. Acad. Sci} {\bf A33}, 264 (1951).		
	
		\bibitem{santos} N.O. Santos, {\em M.N.R.A.S} {\bf 216}, 403 (1985).
		
	
	\bibitem{Liddle:1998xm}
	A.~R.~Liddle and R.~J.~Scherrer,
	Phys. Rev. D \textbf{59} (1999), 023509
	
	
	\bibitem{Uzan:1999ch}
	J.~P.~Uzan,
	{\em Phys. Rev.} D \textbf{59} (1999), 123510
		
		
		
	\bibitem{kumei} G.W. Bluman and S. Kumei, Symmetries and Differential
	Equations, Springer-Verlag, New York, (1989)
	
	\bibitem{sym1} T. Christodoulakis, N. Dimakis and P.A.\ Terzis, J. {\em Phys.\ A:
	Math. Theor.} 47, 095202
	
	\bibitem{sym2} R. Mohanlal, S.D. Maharaj, A.K.\ Tiwari and R. Narain, {\em Gen.\
	Relativ. Gravit.} {bf 48}, 87 (2016)
	
	\bibitem{sym3} M.C. Kweyama, K.S. Govinder and S.D. Maharaj, {\em Class. Quantum
	Grav.} {\bf 28}, 105005 (2011)
	
	\bibitem{sym4} G. Abebe, S.D. Maharaj and K.S. Govinder,{em Eur. Phys. J.} C. {\bf 75}, 496 (2015)
	
	\bibitem{sym5} M.\ Tsamparlis and A. Paliathanasis, Symmetry 10, 233 (2018)
	
	\bibitem{mb1} V.V. Morozov, Classification of six-dimensional nilpotent Lie
	algebras, \textit{Izvestia Vysshikh Uchebn Zavendeni\u{\i} Matematika,} 5,
	161 (1958)
	
	\bibitem{mb2} G.M Mubarakzyanov, Izvestia Vysshikh Uchebn Zavendeni\u{\i}
	Matematika, 32, 114 (1963)
	
	\bibitem{mb3} G.M Mubarakzyanov Izvestia Vysshikh Uchebn Zavendeni\u{\i}
	Matematika, 34, 99 (1963)
	
	\bibitem{mb4} G.M Mubarakzyanov Izvestia Vysshikh Uchebn Zavendeni\u{\i}
	Matematika, 35, 104 (1963)
	
	\bibitem{ars1} M.J. Ablowitz, A. Ramani and H. Segur, \ Lettere al Nuovo
	Cimento 23, 333 (1978)
	
	\bibitem{ars2} M.J. Ablowitz, A. Ramani and H. Segur, J. Math. Phys. 21, 715
	(1980)
	
	\bibitem{ars3} M.J. Ablowitz, A. Ramani and H. Segur, J. Math. Phys. 21,
	1006 (1980)
	
	\bibitem{anl1} A. Paliathanasis and P.G.L.\ Leach, Int. J. Geom. Meth. Mod.
	Phys. 13, 1630009 (2016)	
	
		\bibitem{mg1} S. D. Maharaj and M. Govender, {\em IJMP} {\bf D14}, 667 (2005).
		
		\bibitem{royy} R. Maartens, Causal Thermodynamics in Relativity, arXiv:astro-ph/9609119v1 (1996).	
	
	
	\bibitem{mg2} M. Govender, S. D. Maharaj and R. Maartens,  {\it Class. Quantum Grav.}, {\bf 15}, 323 (1998).
	
	

		\bibitem{mart} J. Martinez, {\it Phys. Rev. D}  {\bf {53}}, 6921 (1996).
		
		\bibitem{wagh1} S. M. Wagh, M. Govender, K. S. Govinder, S. D. Maharaj, P.. S Muktibodh and M. Moodley, {\it Class. Quantum Grav.}, {\bf 18}, 2147 (2001)
		
		\bibitem{rs1} M. Govender, R. S. Bogadi and S. D. Maharaj, {\it Int. J. Mod. Phys.}D {\bf 26}, 1750065 (2017).
		
	\bibitem{hh1} A. Di Prisco, L. Herrera and M. Esculpi, {\em Class. Quantum Grav.},  {\bf 13} 1053
	(1996)
		
		\bibitem{hh2} A. Di Prisco, N. Falc\'on, L. Herrera, M. Esculpi N. O. and Santos N O {\em Gen. Relativ. Gravit.}, {\bf 29} 1391 (1997).
	
	\bibitem{hh3} J. Fort and J. E. and Llebot {\em J. Phys. A: Math. Gen.}, {\bf  29} 3427 (1996).
	
	\bibitem{kesh1} M. Govender and K. S. Govinder, {\it Phys. Lett. A}  {\bf 283}, 71 (2001).
	
	\bibitem{pi} L. Herrera, {\it Phys. Rev. D}, {\bf  101}, 104024 (2020)	
	
\end{thebibliography}
\end{document}